\documentstyle[11pt,psfig]{article}
\makeatletter
\renewcommand{\section}{\@startsection{section}{1}{0in}
	{0.4\baselineskip}{0.1\baselineskip}{\Large\bf}}
\renewcommand{\subsection}{\@startsection{subsection}{2}{0in}
	{0.25\baselineskip}{-\baselineskip}{\large\bf}}
\renewcommand{\subsubsection}{\@startsection{subsubsection}{3}{0in}
	{0.1\baselineskip}{-\baselineskip}{\normalsize\bf}}
\makeatother

\newcommand{\gray}{$\gamma$-ray}
\newcommand{\grays}{$\gamma$-rays}
\begin{document}

%
\thispagestyle{myheadings}
%
\markright{OG 2.1.05}
\begin{center}
%
{\LARGE \bf GLAST and the Extragalactic Gamma Ray Background}
\end{center}

\begin{center}
%
%
{\bf F.W. Stecker$^{1}$ and M.H. Salamon$^{2}$}\\
{\it $^{1}$Laboratory for High Energy Astrophysics, NASA/GSFC, Greenbelt, MD 20771, USA\\
$^{2}$Physics Department, University of Utah, Salt Lake City, UT 84112-0830, USA}
\end{center}

\begin{center}
{\large \bf Abstract\\}
\end{center}
\vspace{-0.5ex}
%
%
We show that GLAST should be able to support or rule out the unresolved-blazar
hypothesis for the origin of the diffuse, extragalactic gamma ray background.
%

\vspace{1ex}

%
%
\section{Introduction:}
\label{intro.sec}

Perhaps the most promising model proposed for the origin of the $\sim$GeV extragalactic \gray\ 
background (EGRB), first detected by SAS-2 [Fichtel, Simpson, \& Thompson 1978] 
and later confirmed by EGRET,
is that it is the collective emission of an isotropic distribution of faint,
unresolved blazars [Stecker \& Salamon 1996 and references therein].  EGRET has determined that the EGRB spectrum is consistent with a single
power-law,
\begin{equation}
\frac{dN_{\gamma}}{dE}=(7.32\pm 0.34)\times 10^{-6}\left(\frac{E}{0.451 {\rm GeV}}\right)
^{-2.10\pm0.03} {\rm cm}^{-2}{\rm s}^{-1}{\rm sr}^{-1}{\rm GeV}^{-1}
\end{equation}
between 0.1 and $\sim 50$ GeV (statistics limited) [Sreekumar et al. 1998].  
Because GLAST has a roughly 
energy-independent area of $10^{4}$ cm$^{2}$ above 0.1 GeV (compared to EGRET's $10^{3}$
cm$^{2}$ at 1 GeV and $10^{2}$ cm$^{2}$ at 100 GeV), with an estimated point source
sensitivity (PSS) nearly two orders of magnitude lower than EGRET's, GLAST 
will be able
to (a) detect something on the order of $10^{2}$ times more blazars than EGRET, and (b) measure the EGRB
spectrum to $>1$ TeV (assuming the EGRET power law spectrum).  These two capabilities will
enable GLAST to either strongly support or reject the unresolved-blazar hypothesis for
the origin of the EGRB.

\section{The Unresolved Blazar Model:}
\label{model.sec}

To determine the collective output of all \gray\ blazars, one can use the observed
EGRET distribution of \gray\ luminosities and extrapolate to obtain a ``direct'' \gray\
luminosity funtion (LF) $f_{\gamma}(l_{\gamma},z)$ [Chiang, et al. 1995 ] ($f_{\gamma}$ units being
sources/co-moving volume-differential luminosity), where $l_{\gamma}$ is the differential
luminosity (s$^{-1}$) at a (source) fiducial photon energy $E_{f}$, and we assume power
law spectra for all sources, $l(E)=l_{\gamma}(E/E_{f})^{-\alpha}$.  Alternatively, one
can make use of much larger catalogs at other wavelengths, and assume some relationship
between the source luminosities at the catalog wavelength and the GeV region 
[Padovani et al. 1993; Stecker, Salamon, \& Malkan 1993].
Both methods are fraught with potential errors.  In the former method, only the ``tip
of the iceberg'' of the \gray\ LF has been observed by EGRET, and extrapolation to
lower luminosities that fall below EGRET's PSS (for any reasonable source distance)
involves some level of assumption.  In the latter method, the assumption of a linear
relation between the luminosities of a source at disparate wavelengths is by no means
well established [Padovani, et al. 1993; Muecke, et al 1996, Mattox, et al. 1997].

We used the latter method in the past [Stecker and Salamon 1996] 
to estimate the contribution of unresolved
blazars to the EGRB, and found that up to 100\% of EGRET's EGRB can be accounted for.  This
model assumes a linear relationship between the differential \gray\ luminosity $l_{\gamma}$
at $E_{f}=0.1$ GeV and the differential radio luminosity $l_{r}$ at 2.7 GHz for all sources,
{\it viz.,} $l_{\gamma}=\kappa l_{r}$, where $\kappa$ is a constant.  One can then use the
measured radio LF $f_{r}(l_{r},z)$ for blazars (primarily for flat spectrum radio quasars)
[Dunlop \& Peacock 1990] to calculate the collective \gray\ output of all blazars.

The simplified elements of the calculation follow (the details can be found in 
Stecker and Salamon 1996).
First, the \gray\ and radio LFs are related by 
$f_{\gamma}(l_{\gamma},z)=\kappa^{-1}f_{r}(\kappa^{-1}l_{\gamma},z)$.
The number of
sources ${\cal N}$ detected by a detector is a function of the detector's 
point source sensitivity (PSS) at the
fiducial energy $E_{f}$, $[F(E_{f})]_{\rm min}$, where the integral \gray\ number flux
$F$ is related to $l_{\gamma}$ by
\begin{equation} \label{1.eq}
F(E)=l_{\gamma}(E/E_{f})^{-\alpha}/4\pi\alpha (1+\alpha)^{\alpha +1}R_{0}^{2}r^{2},
\end{equation}
where $R_{0}r$ is the luminosity distance to the source.  The number of sources at
redshift $z$ seen at Earth with an integral flux $F(E_{f})$ is given by
\begin{equation}
\frac{d{\cal N}}{dF(E_{f})}\Delta F(E_{f})=
\int 4\pi R_{0}^{3}r^{2}\, dr\, f_{\gamma}(l_{\gamma},z(r))\Delta l_{\gamma},
\end{equation}
where $l_{\gamma}$ in the integrand depends on $z(r)$ and $F(E_{f})$ from 
Eq.\ref{1.eq}.  Figure 1 shows our 1996 calculations for the number of sources
versus flux, compared to the EGRET detections.  The cutoff at
$\sim 10^{-7}{\rm cm}^{-2}{\rm s}^{-1}$ for $E_{f}=0.1$ GeV, their quoted PSS,
is evident.

\begin{figure}[ht]
\psfig{file=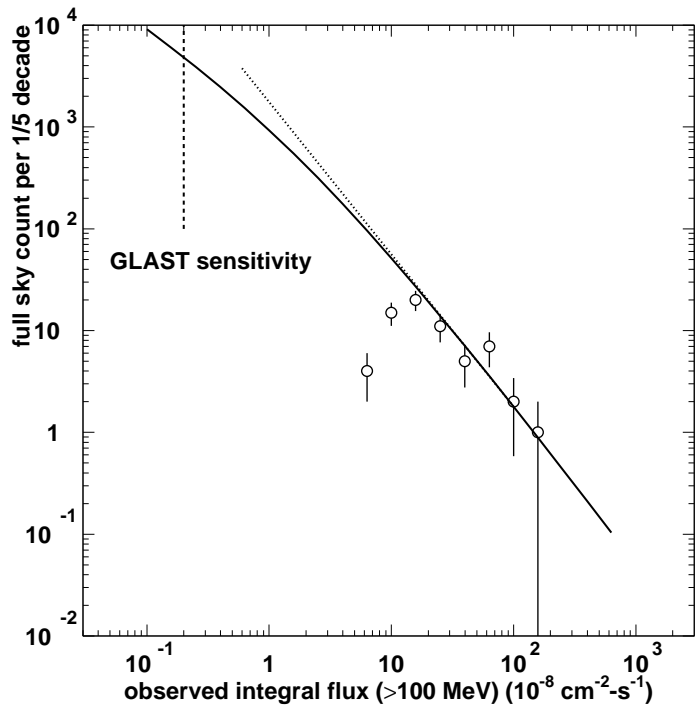,width=4.0in}
\vspace{-0.1cm}
\caption{Source number count per one-fifth decade of integral flux at Earth.  The
straight dotted line is the Euclidean relation ${\cal N}(>F)\propto F^{-3/2}$ for
homogeneous distribution of sources.  The open circles represent the EGRET blazar
detections, and the solid line is the model prediction.}
\end{figure}
\vspace{1.0cm}
To calculate the EGRB, we integrate over all sources {\it not} detectable by the
telescope to obtain the differential number flux of EGRB photons at an {\it observed}
energy $E_{0}$:

\begin{equation}
\frac{dN_{\gamma}}{dE}(E_{0})=\int 4\pi R_{0}^{3}r^{2}\, dr
\int d\alpha \, p(\alpha)
\int_{l_{min}}^{l_{max}}\frac{dF}{dE}(E_{0}(1+z))f_{\gamma}(l_{\gamma},z)
e^{-\tau(E_{0},z)}\, dl_{\gamma}.
\end{equation}
This expression includes an integration over the probability distribution of
spectral indices $\alpha$ (based on the 2nd EGRET Catalog [Thompson et al. 1995]).  
There
is also an important attenuation factor in this expression, due to the loss of
\grays\ as they propagate through the intergalactic medium and interact with 
cosmic UV, optical, and IR background photons to produce $e^{\pm}$ pairs.
Although there are significant uncertainties in estimates of the $z$-dependence
of the soft photon background [Salamon \& Stecker 1998], 
a qualitative feature cannot be avoided:
If a substantial fraction of the EGRB is from high-$z$ flat spectrum radio quasars, a 
steepening in the spectrum should be seen at energies above 20 GeV.
Figure 2 shows the calculated EGRB spectrum (based on EGRET's PSS) compared to
EGRET data.  The ``bump'' in the spectrum above 10 GeV is due to there 
being a finite width in the assumed power-law spectral index distribution of blazars, which produces a 
summed unabsorbed spectrum with a positive second derivative [Stecker \& Salamon 1996].
Although the
match between the model's spectrum and the recent EGRET EGRB data [Sreekumar et al. 1998]
does not appear to be very good, one must examine it the light of the assumptions and uncertainties.
The location and amplitude of the ``bump'' depend upon the amount of high redshift absorption, which, in turn, depend upon
uncertain UV and optical soft photon background densities [Salamon \& Stecker 1998]. The ``bump'' also depends upon the poorly known
blazar
spectral index distribution, which we have assumed to independent of quiescent-state
luminosity. Finally, we not that the assumed power-law form for high redshift quasars above 10 GeV energy is highly questionable in light of the predictions of the Compton models for \gray\ emission from these sources; in fact their spectra are predicted to steepen at these energies.
We also note that EGRET's EGRB error bars are systematic, not statistical and that Sreekumar et al. [1998] have stated that the model cannot be ruled out based on the EGRET data.

\section{GLAST and the EGRB:}
\label{glast.sec}

Figure 1 shows that $\cal O$$(10^{3})$ blazars should be detectable by GLAST,
assuming it achieves a PSS of $\sim 2\times 10^{-9}$ cm$^{-2}$s$^{-1}$, and Figure 2
shows that this should reduce the EGRB by a factor of $\sim 2$ for $E>1$ GeV.  These
predictions are arrived at by replacing EGRET's PSS with GLAST's in our 1996 model.

\begin{figure}[ht]
\psfig{file=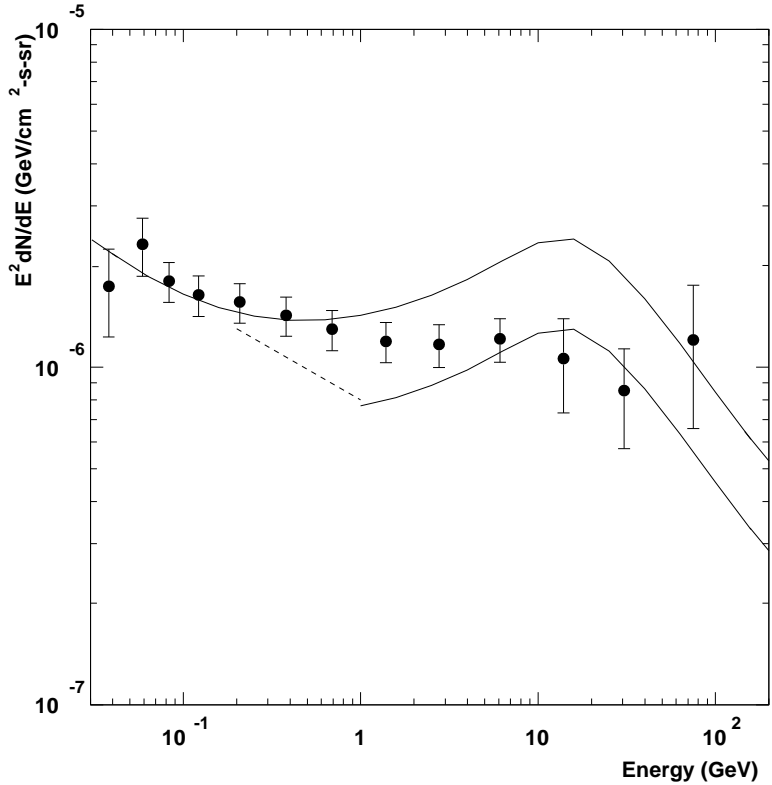,width=4.0in}
\vspace{-0.1cm}
\caption{The upper and lower solid curves correspond to model predictions for
the EGRB spectrum given EGRET's and GLAST's PSS (respectively).  The data points
are from the latest EGRET EGRB analysis [Sreekumar et al. 1998].  The dashed line
segment in the lower curve corresponds to the energy interval where increasing
source confusion, owing to worsening angular resolution at lower energies, prevents
the discernment of point sources, which then contribute to the EGRB.}
\vspace{1.0cm}
\end{figure}

Figure 2 shows our predicted reduction in the EGRB from unresoved blazars to be observed by GLAST compared with that observed by EGRET.
The dashed line in Figure 2 represents the energy interval in which it is difficult
to determine GLAST's EGRB spectrum, due to the effect of worsening angular resolution
with decreasing energy.  For a lower PSS level, the number of
detectable sources is larger.  However, the angular resolution limits the number of
distinct sources one can identify, crudely given by
${\cal N}(E)\approx 4\pi/\pi\sigma_{\theta}^{2}(E)$, where $\sigma_{\theta}(E)$ is
the energy-dependent angular resolution.  When ${\cal N}(>F_{min})>{\cal N}(E)$,
individual sources are no longer separable from the EGRB.  With GLAST's proposed angular 
resolution function (GLAST Science Document), the equality ${\cal N}(>F_{min})={\cal N}(E)$ occurs
at $E$ somwhat less than 1 GeV; thus for $E<1$ GeV there is not as much a reduction in the level
of the EGRB (compared to EGRET) as for energies $E>1 $ GeV.

We conclude that GLAST can test the unresolved blazar background model in three ways: (1) GLAST should see roughly 2 orders of magnitude more blazars than EGRET because of its ability to detect the fainter blazars which contribute to the EGRB in our model. Moreover, GLAST can test our assumption of an average linear relation between \gray\ and radio flux and our assumed redshift distribution of blazars by testing the details of our source count versus flux prediction. (2) With GLAST's improved PSS leading to more blazars being resolved out, fewer unresolved blazars will be left to
contribute to the EGRB, thus reducing the level of the measured EGRB compared to
EGRET's.
(3) GLAST's much greater aperture at 100 GeV will allow a determination
of whether or not a steepening exists in the EGRB, since
the number of EGRB \grays\
recorded by GLAST above 100 GeV will be of order $10^{3}$ to $10^{4}$, assuming a
continuation of the EGRET power law spectrum. However, this last test must be qualified because of the unknown amount of steepening in the high reshift quasar spectra above 10 GeV which can mimic an absorption effect.

%
%
\vspace{1ex}
\begin{center}
{\Large\bf References}
\end{center}
%
Chiang, J. et al. 1995, ApJ 452, 156.\\
Dunlop, J.S. and Peacock, J.A. 1990, MNRAS 247, 19.\\
Fichtel, C.E., Simpson, G., \& Thompson, D.J. 1978, ApJ 222, 833.\\
Mattox, J.R., et al. 1997, ApJ 481, 95.\\
Muecke, A., et al. 1996, A\&AS 120, 541.\\
Padovani, P. et al. 1993, MNRAS 260, L21.\\
Salamon, M.H. and Stecker, F.W. 1998, ApJ 493, 547.\\
Sreekumar, P. et al. 1998, Ap.J. 494, 523.\\
Stecker, F.W., Salamon, M.H., \& Malkan, M.A. 1993, ApJ 410, L71.\\
Stecker, F.W. and Salamon, M.H. 1996, ApJ 464, 600.\\
Thompson, D.J. et al. 1995, ApJS 101, 259.\\

\end{document}